\documentclass[prl,twocolumn,showpacs,preprintnumbers,amsmath,amssymb]{revtex4}
\usepackage{graphicx}
\usepackage{dcolumn}
\usepackage{bm}
\usepackage{color}

\begin{document}

\title{Photoelectron circular dichroism with two overlapping laser pulses of carrier frequencies $\omega$ and $2\omega$ linearly polarized in two mutually-orthogonal directions}

\author{\firstname{Philipp~V.} \surname{Demekhin}}\email{demekhin@physik.uni-kassel.de}
\affiliation{Institute of Physics and CINSaT, University of Kassel, Heinrich-Plett-Str.~40, 34132 Kassel, Germany}

\author{\firstname{Anton~N.} \surname{Artemyev}}
\affiliation{Institute of Physics and CINSaT, University of Kassel, Heinrich-Plett-Str.~40, 34132 Kassel, Germany}

\author{\firstname{Alexander} \surname{Kastner}}
\affiliation{Institute of Physics and CINSaT, University of Kassel, Heinrich-Plett-Str.~40, 34132 Kassel, Germany}

\author{\firstname{Thomas} \surname{Baumert}}\email{baumert@physik.uni-kassel.de}
\affiliation{Institute of Physics and CINSaT, University of Kassel, Heinrich-Plett-Str.~40, 34132 Kassel, Germany}

\date{\today}

\begin{abstract}
Using a model methane-like chiral system, we theoretically demonstrate a possibility to access photoelectron circular dichroism (PECD) by a single experiment  with two overlapping laser pulses of carrier frequencies $\omega$ and $2\omega$, which are linearly polarized in  two mutually-orthogonal  directions. Depending on the relative phase, the resulting electric field can be tailored to have two different rotational directions in the upper and lower hemispheres along the polarization of the $\omega$-pulse. We predict a strong forward/backward asymmetry in the emission of photoelectrons from randomly oriented samples, which has an opposite sign in the upper and lower hemispheres. The predicted PECD effect is phase- and enantiomer-sensitive,  providing new insight in this fascinating fundamental phenomenon. The effect can be optimized by varying relative intensities of the pulses.
\end{abstract}


\maketitle

In 1976, it was predicted theoretically that photoionization of chiral molecules is sensitive to the helicity of ionizing light \cite{Ritchie}. The effect emerges already in the electric-dipole approximation and manifests itself as a forward/backward asymmetry in the emission of photoelectrons from randomly oriented chiral molecules in the gas phase. It took about 25 years to verify these theoretical predictions experimentally \cite{CDwf01,CDwf02}. Later on, this enantiomer- and helicity-selective effect was termed as photoelectron circular dichroism (PECD) \cite{Define}. During the last decade, one-photon ionization of chiral molecules by circularly polarized radiation, and the emerging PECD effects, were studied in the numerous experimental and theoretical works  (see, e.g., review articles \cite{REV1,REV2,REV3}).

Recently \cite{Lux12AngChm,Lehmann13jcp}, a similar PECD effect has been observed in the multiphoton ionization of chiral molecules by circularly polarized  laser pulses. Since then, extensive experimental studies provided many important details on the multiphoton PECD, like, e.g., on its dependence on the pulse intensity and ellipticity \cite{Lux15CPC,Lux16ATI,Beaulieu16NJP}, on the enantiomeric excess of the target \cite{Kastner16ee,Miles17ee}, or on the intermediate electronic states involved in different multiphoton ionization schemes \cite{Rafiee16wl,Kastner17wl}. In addition, an impact of the nuclear and electron  dynamics during multiphoton ionization of chiral molecules
have been demonstrated by pump-probe experiments \cite{Beaulieu16td,Comby16td,Beaulieu17as,Beaulieu18PXCD}. Theoretical approaches to describe the multiphoton PECD range from time-independent perturbative ab-initio calculation of the two-photon absorption followed by one-photon ionization in hydrogenic continuum spectrum \cite{Goetz17} to nonperturbative time-dependent methods \cite{TDSC1,TDSC2}.

Employing overlapping bichromatic laser pulses is a particularly important example of coherent control schemes \cite{CoCo1,CoCo2}. Here, interference between the $n$-photon route of one of the fields and the $m$-photon route of the other field can be controlled through the relative phase. According to the selection rules for multiphoton transitions in atoms, there are two control scenarios \cite{CoCo1}. If $n$ and $m$ are both odd or even  numbers, the integral and differential cross section can be controlled. In the case where $n$ and $m$ are either odd or even, only differential cross section, i.e., photoemission in different angles, can be controlled. The latter effect can also be understood in terms of symmetry-breaking of the total electric field of two pulses. The simplest realization of such an interference is the utilization of integer multiples of frequency- components of a fundamental laser field \cite{Muller92,Yin92,Schumacher94,Yin95,Wang01,Yamazaki07,Vortex1,Vortex2,Vortex3,XUV}.

\begin{figure}
\includegraphics[scale=0.38]{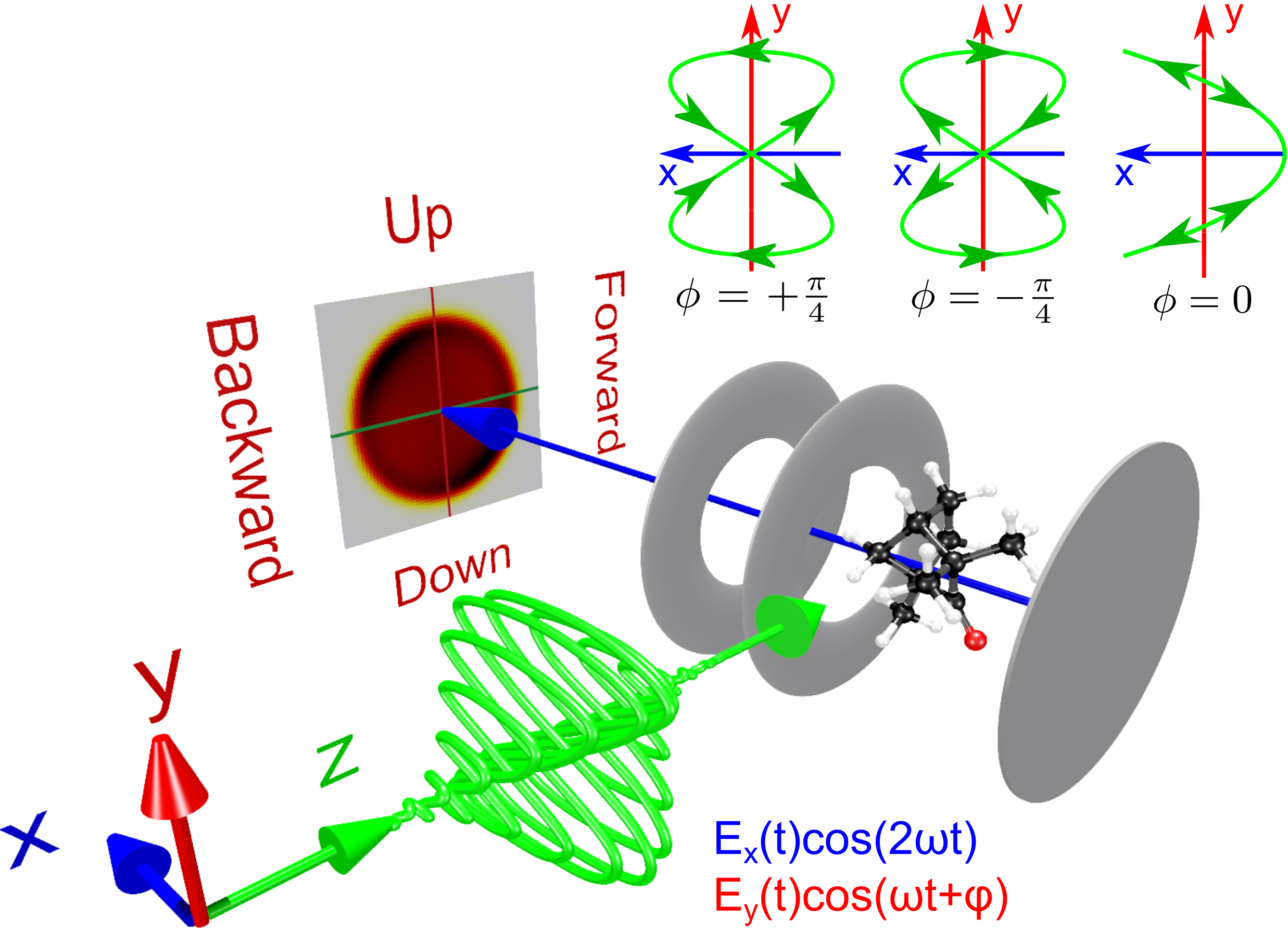}
\caption{Scheme of the experiment with bichromatic laser pulses (\ref{field}). The pulses propagate along $z$-axis. For the relative phase $\phi=+ \frac{\pi}{4}$, the resulting electric field rotates counter-clockwise in the upper and clockwise in the lower parts of $xy$-plane from the point of view of the emitter. During measurements, photoelectrons are projected on the $yz$-plane of, e.g., a VMI detector. Thereby, photoelectrons released by the field pointing in the upper/lower hemispheres are projected onto the upper/lower parts of the detector (labeled as `Up/Down'). For a chiral target in the gas phase, it is expected that velocity map images recorded by the detector will exhibit an opposite sign of the forward/backward asymmetry (of the PECD) on its upper and lower parts.} \label{fig1}
\end{figure}

Polarization- and phase-locked bichromatic pulses have been successively used to control angular emission distribution of photoproducts in molecules \cite{Gong14,Wu13} and photocurrent in semiconductors \cite{Dupont95,Hache97} and on metal surfaces \cite{Gudde07}. Here we demonstrate a possibility to utilize bichromatic pulses to control PECD. For this purpose, we propose an experiment on the photoionization of chiral molecules  by two temporally-overlapping laser pulses with carrier frequencies of $\omega$ and $2\omega$, which are polarized linearly along two perpendicular directions. Depending on the relative phase $\phi$, the resulting electric field
\begin{equation}
\mathcal{\vec{E}}(t)= \hat{e}_x \mathcal{E}_x \,g(t) \cos(2\omega t) + \hat{e}_y \mathcal{E}_y \,g(t) \cos(\omega t +\phi)
\label{field}
\end{equation}
can have a strong asymmetry in the $xy$-plane. Here, $\mathcal{E}_x$ and $\mathcal{E}_y$ are field amplitudes, and $g(t)$ is the time-envelope of the pulses. For instance for $\phi=0$ or $\phi=\pm\frac{\pi}{2}$, the field $\mathcal{\vec{E}}(t)$  has a `horseshoe' form, which is asymmetric along the $x$-axis. For the case of $\phi=\pm \frac{\pi}{4}$, the resulting electric field has a `butterfly' form, which is oriented along the $y$-axis (see upper right part of Fig.~\ref{fig1}). In the latter case, the resulting field is mirrored with respect to $x$-axis, and it exhibits two different rotational directions for positive and negative values of $y$.

The geometry of the presently proposed experiment is illustrated in Fig.~\ref{fig1}. In the upper hemisphere ($y>0$), the bichromatic electric field (\ref{field}) points upwards and has one of the rotational directions. In the lower hemisphere ($y<0$), it points downwards and has an opposite rotational direction. During photoionization, photoelectrons are released along the electric field in the upper/lower hemispheres and will subsequently be projected on the respective `Up/Down' parts of the velocity map imaging (VMI, \cite{VMI1,VMI2}) detector placed in $yz$-plane. Such a polarization state allows for a single and simultaneous measurement with two different rotational directions in the upper and lower hemispheres. A chiral target acts as a `gearbox' \cite{REV1,TIA17} which transforms rotational motion of the electric field in the translation motion of the photoelectron  along the pulse propagation direction ($z$-axis). One thus can expect a different sign of the forward/backward ($z>0/z<0$) asymmetry in the photoelectron angular distribution images on the upper/lower ($y>0/y<0$) parts of the detector.
Such an experiment can be performed by combining the fundamental and its second harmonic driving two-red-photon vs. one-blue-photon ionization of a chiral molecule.

In order to verify this hypothesis, we simulated the proposed experiment theoretically. Calculations were carried out by the time-dependent Single Center (TDSC) method and code \cite{TDSC1,TDSC2}. It consists in the propagation of the wave packet of a single-active-electron, which is driven in the potential of a chiral ion by an intense short laser pulse. The TDSC method accounts for the light-matter interaction nonperturbatively via the numerical solution of the time-dependent Schr\"{o}dinger equation. Details of the method can be found in Ref.~\cite{TDSC1}. There, it was applied to study PECD in the one-photon ionization and two-photon above-threshold ionization of a model methane-like chiral system by short intense high-frequency laser pulses.

\begin{figure}
\includegraphics[scale=0.62]{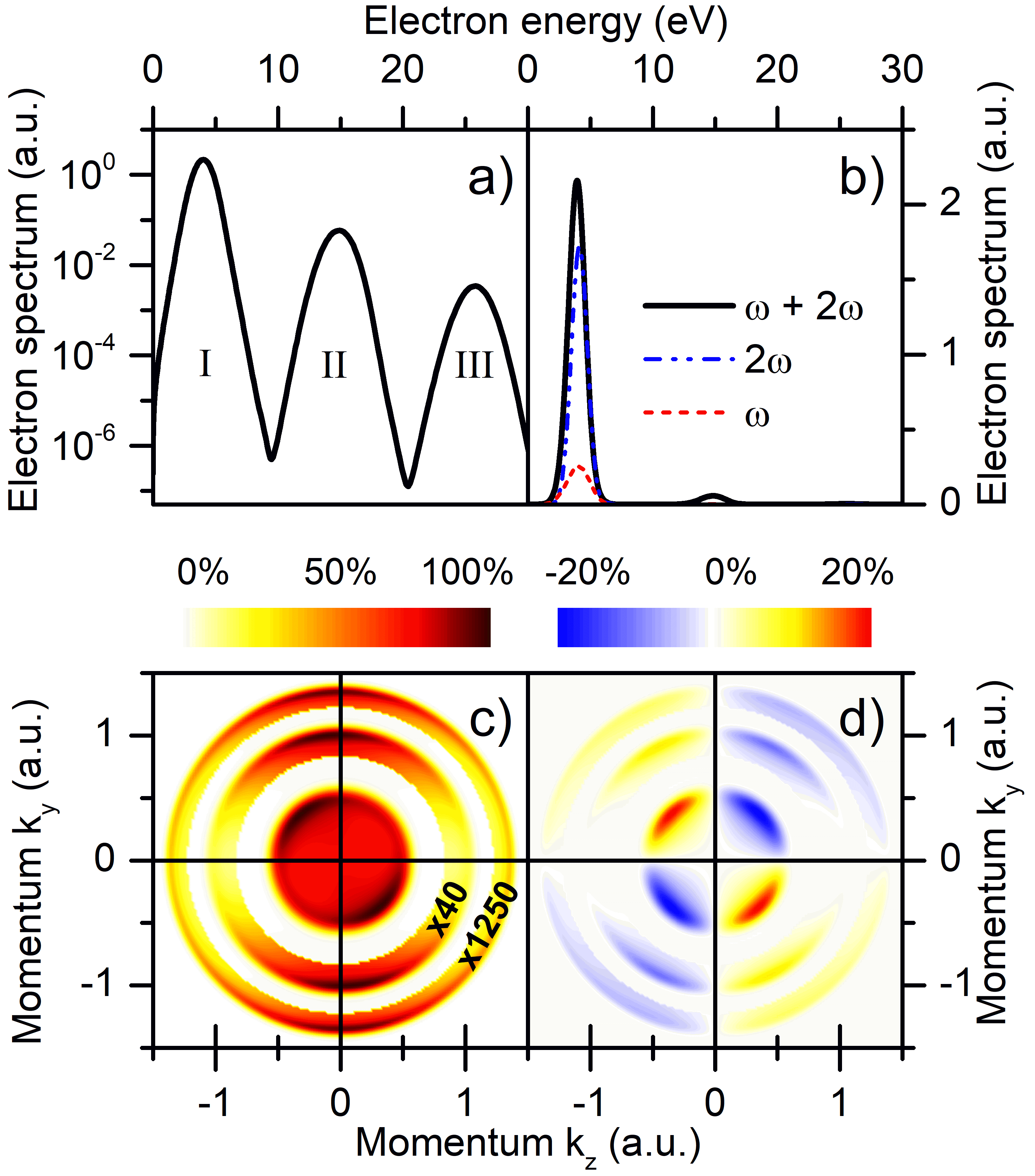}
\caption{{Panel a)}: Total photoelectron spectrum computed for the model chiral system exposed to the bichromatic field (\ref{field}) with $\phi=+\frac{\pi}{4}$ `butterfly' form. Note the logarithmic scale of the vertical axis. Peak~I is produced by the absorption of either one blue photon or two red photons; peak~II -- by either one blue and one red photon or by three red photons; peak III -- by either two blue photons or one blue and two red photons or four red photons. {Panel b)}: Individual contributions to the total spectrum (solid curve, the same as in panel a)) from each of the linearly polarized pulses with carrier frequencies $2\omega$ and $\omega$ (see legend). Panel c): Projection of the photoelectron angular distribution on the detector $yz$-plane. Maximum of the intensity is set to 100\%. Note that two outer rings are shown on the enhanced scales, as indicated by the factors $\times 40$ and $\times 1250$ in this panel. Panel d): photoelectron circular dichroism in percent of the total intensity of the signal. Note from panels c) and d) that the forward/backward asymmetry in the emission of  photoelectrons (along the $k_z$-axis) has different sign in the upper and lower hemispheres (along the $k_y$-axis).} \label{fig2}
\end{figure}

\begin{figure*}
\includegraphics[scale=0.62]{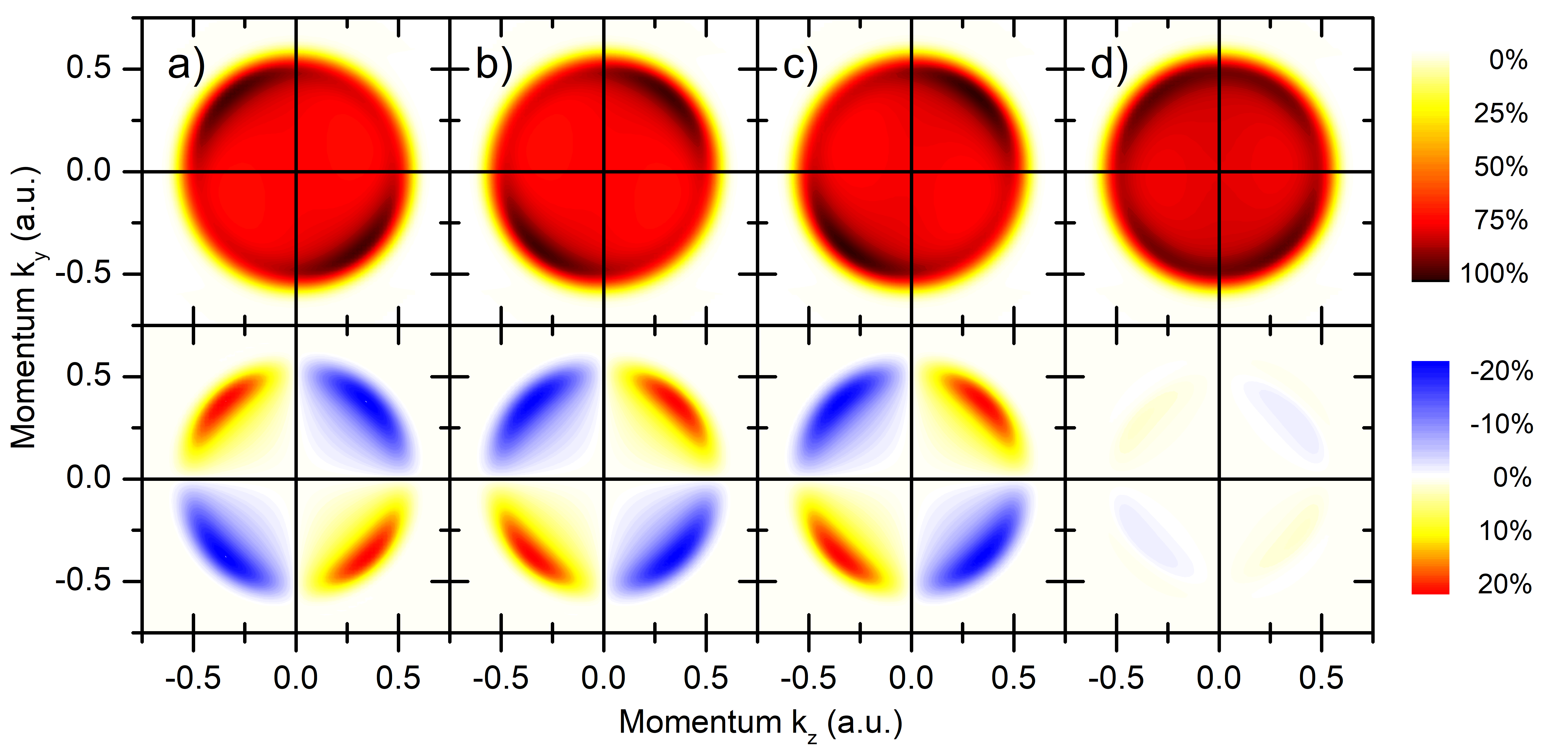}
\caption{Projections of the photoelectron angular distributions on the detector yz-plane (upper panels) and the respective PECDs (lower panels) computed for the model chiral system and different phases in Eq.~(\ref{field}): (a) $\phi=+\frac{\pi}{4}$ `butterfly' form, (b)  $\phi=-\frac{\pi}{4}$ `butterfly' form, and (d) $\phi=0$  `horseshoe' form. Only the contribution from peak~I is shown. Results computed for another enantiomer of the system and phase $\phi=+\frac{\pi}{4}$ (`butterfly' form) are shown in panel (c). Switching the phase $\phi$ from $+\frac{\pi}{4}$ to $-\frac{\pi}{4}$ is thus equivalent to switching enantiomers, which confirms a chiral origin of the effect. Note that color-scales of all upper panels (and separately of all lower panels) are identical. An asymmetry of about 1\% seen in panel (d) can be considered as the accuracy of the present integration over different molecular orientations.} \label{fig3}
\end{figure*}

In the present work we use the  methane-like model chiral system from Ref.~\cite{TDSC1} (see Fig.~1 there) and similar high-frequency laser pulses. In particular, we utilize short gaussian-shaped pulses of $g(t)=\exp(-(t-t_0)^2/\tau^2)$ with $\tau=1$~fs. The carrier frequency of the `red' pulse is set to $\omega=11.05$~eV, such that absorption of one `blue' photon of energy $2\omega=22.1$~eV results in the ionization of the system with the ionization potential of 18.3~eV \cite{TDSC1}. We, thus, realize the two-red-photon vs. one-blue-photon ionization scheme by the $\omega$ and $2\omega$ pulses. In Ref.~\cite{TDSC1}, it was demonstrated that one-photon ionization spectrum of this model system by circularly polarized $2\omega=22.1$~eV pulse exhibits a sizable PECD effect of about 15\%.

Owing to the chosen high frequencies, these short pulses support about 10 and 20 optical cycles for the $\omega$ and $2\omega$ components, respectively. Therefore, any asymmetry effects due to the  carrier-envelope phase can be neglected. The field (\ref{field}) possesses no axial symmetry along the pulse propagation direction. Therefore, in order to simulate the gas phase experiment, ionization spectra computed at different molecular orientations should be averaged over \emph{all} rotational Euler angles ($\alpha, \beta,\gamma$) for the transformation of the molecular to the laboratory frame. The orientation intervals $\alpha,\gamma\in[0,2\pi)$ and $\beta\in[0,\pi]$ were covered with the steps of $\Delta \alpha= \Delta \beta=\Delta \gamma =0.1\,\pi$. Finally, owing to the same reason, the photoelectron angular distribution images recorded by, e.g., a VIM detector cannot be Abel-inverted. Therefore, we theoretically simulate these images by projecting (by integrating over $x$-coordinate) the computed three-dimensional angular distributions on the detector $yz$-plane (see Fig.~\ref{fig1}).

The main set of the present calculations was performed for the equal peak intensities of two pulses $I_x=I_y=10^{14}$~W/cm$^2$ (see below). An overview of the total electron spectrum, computed for $\phi=+\frac{\pi}{4}$ and integrated over all molecular orientations, is shown in Fig.~\ref{fig2}a. It exhibits three main features. The strongest threshold peak~I at about $\varepsilon=3.8$~eV  is produced by the competition of  two-photon ionization by the red pulse and  one-photon ionization by the blue pulse. At the chosen pulse intensities, the individual contribution to this peak~I from the red pulse becomes noticeable compared to the contribution from the blue pulse (see Fig.~\ref{fig2}b). The weak features~II and III in Fig.~\ref{fig2}a represent electrons released by the above-threshold ionization processes. Peak~II at about $\varepsilon=14.85$~eV is produced either by the absorption of one blue and one red photon or by three red photons. The very weak peak~III at $\varepsilon=25.9$~eV is due to the absorption of two blue photons, one blue and two red photons, or four red photons.

Figure~\ref{fig2}c depicts the projection of the three-dimensional photoelectron angular distribution on the detector $yz$-plane. One can see that the computed image exhibits a sizable forward/backward asymmetry (along the pulse propagation direction, $k_z$-axis). The asymmetry is present in all peaks I, II, and III. Very important, this asymmetry is opposite for positive and negative values of  $k_y$-momentum (i.e., in the upper and lower hemispheres). In order to set this effect on a quantitative scale, we introduce PECD as the difference between the two signals $I(k_y,k_z)-I(-k_y,k_z)$. Figure~\ref{fig2}d illustrates the effect in percent of the maximal intensity. One can see that computed PECD has a different sign for  $k_y>0$ and $k_y<0$. The peak I exhibits the strongest asymmetry of about 20\%, while the above-threshold ionization peaks II and III show a smaller PECD of 11\% and 6\%, respectively.

In order to confirm the chiral origin of this effect, we performed a set of calculations for  different relative phases and for another enantiomer of the model chiral system. Results of these calculations are summarized in Fig.~\ref{fig3}, which depicts the detector images and the respective PECDs  for the threshold peak I. The data in Figs.~\ref{fig3}a, obtained for $\phi=+\frac{\pi}{4}$, are the same as those in Figs.~\ref{fig2}c and 2d.  Figure~\ref{fig3}b demonstrates that switching the rotational directions of the field to the opposite by changing the phase to $\phi=-\frac{\pi}{4}$ results in the opposite effect of the same size. In addition, switching the enantiomer but keeping the phase $\phi=+\frac{\pi}{4}$ coincides with keeping the enantiomer and switching the phase to $\phi=-\frac{\pi}{4}$ (compare Figs.~\ref{fig3}b and \ref{fig3}c).

\begin{figure}
\includegraphics[scale=0.62]{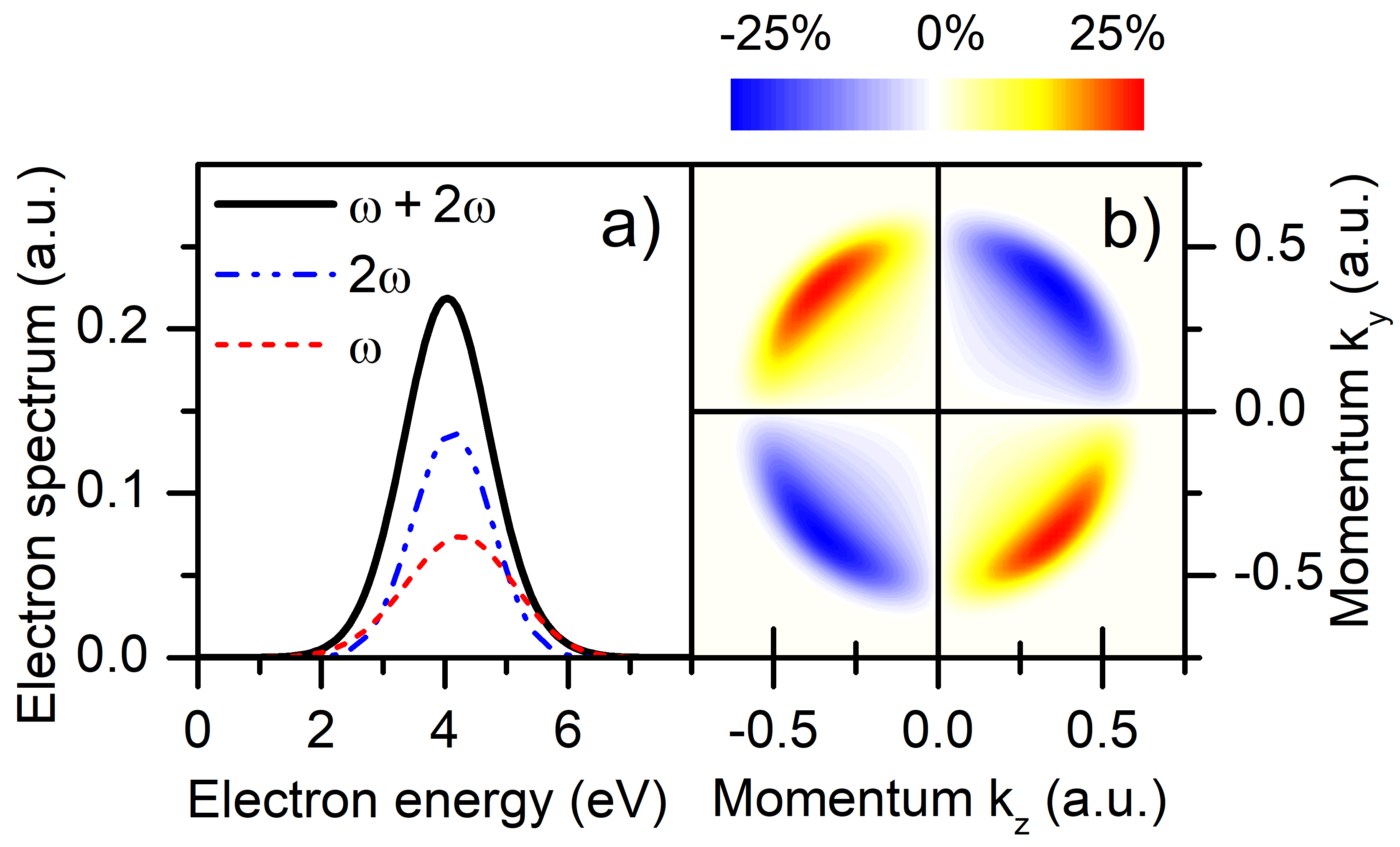}
\caption{Computational results obtained for the case of nonequal field amplitudes $\mathcal{E}_x$ and $\mathcal{E}_y$ (see text). {Panel a)}: Individual contributions to the total spectrum from each of the pulses with carrier frequencies $2\omega$ and $\omega$ (see legend). Panel b): PECD in percent of the total intensity of the signal. The observed maximal asymmetry of about 25\% is larger than that of 20\% seen from Fig.~\ref{fig3}.} \label{fig4}
\end{figure}

For the relative phase $\phi=0$, the resulting electric field (\ref{field}) possesses two different rotational directions for equal periods of time in each of the hemispheres  (see upper right part of Fig.~\ref{fig1}). Such a `horseshoe'-shaped field, thus, exhibits no preferable rotational direction in the upper or lower hemispheres. As is evident from Figs.~\ref{fig3}d, the detector image computed for $\phi=0$ is almost forward/backward and up/down symmetric. A very weak asymmetry seen for $\phi=0$ is within the error of the present integration over molecular orientations due to finite intervals $\Delta \alpha$, $\Delta \beta$, and $\Delta \gamma$. Finally, the three-dimensional photoelectron angular distribution computed for $\phi=0$ is strongly asymmetric along the $x$-axis (not shown here), owing to the respective asymmetry of the electric field (\ref{field}).

As a final point, we demonstrate that the maximal chiral asymmetry, achieved at $\phi=\pm\frac{\pi}{4}$, can be optimized by varying the field amplitudes of two pulses. Indeed, the effect demonstrated here depends not only on the asymmetry of the total electric field, but also on intrinsic electronic properties of a chiral target, i.e., on the respective transition amplitudes for the one-photon and two-photon ionization pathways. In the previous example, the peak intensities of two pulses were equivalent, and the relative contribution of the two-photon ionization was rather small compared to the one-photon ionization (see Fig.~\ref{fig2}b). What happens if the two contributions will be made comparable? This can be achieved by enhancing the $\mathcal{E}_y$-component of the field (\ref{field}).

Figure~\ref{fig4} depicts computational results obtained for the peak intensities $I_x=10^{13}$ and $I_y=4\times10^{13}$~W/cm$^2$ and relative phase $\phi=+\frac{\pi}{4}$. Thereby, $\mathcal{E}_y$-component is two times larger than $\mathcal{E}_x$, and the resulting `butterfly' field is stretched along the $y$-axis. Now, the individual contributions to the total spectrum from the one-photon and two-photon ionization processes are made comparable (see Fig.~\ref{fig4}a). The respective PECD computed for the threshold photoelectron peak~I is shown in Fig.~\ref{fig4}b. The chiral effect in this figure is very similar to that obtained for equal peak intensities of two pulses (see Fig.~\ref{fig3}a). However, PECD in Fig.~\ref{fig4}b is larger than that in Fig.~\ref{fig3}a. This may provide a higher sensitivity in the enantiomeric excess determination \cite{Kastner16ee,Miles17ee}.

\begin{figure}
\includegraphics[scale=0.57]{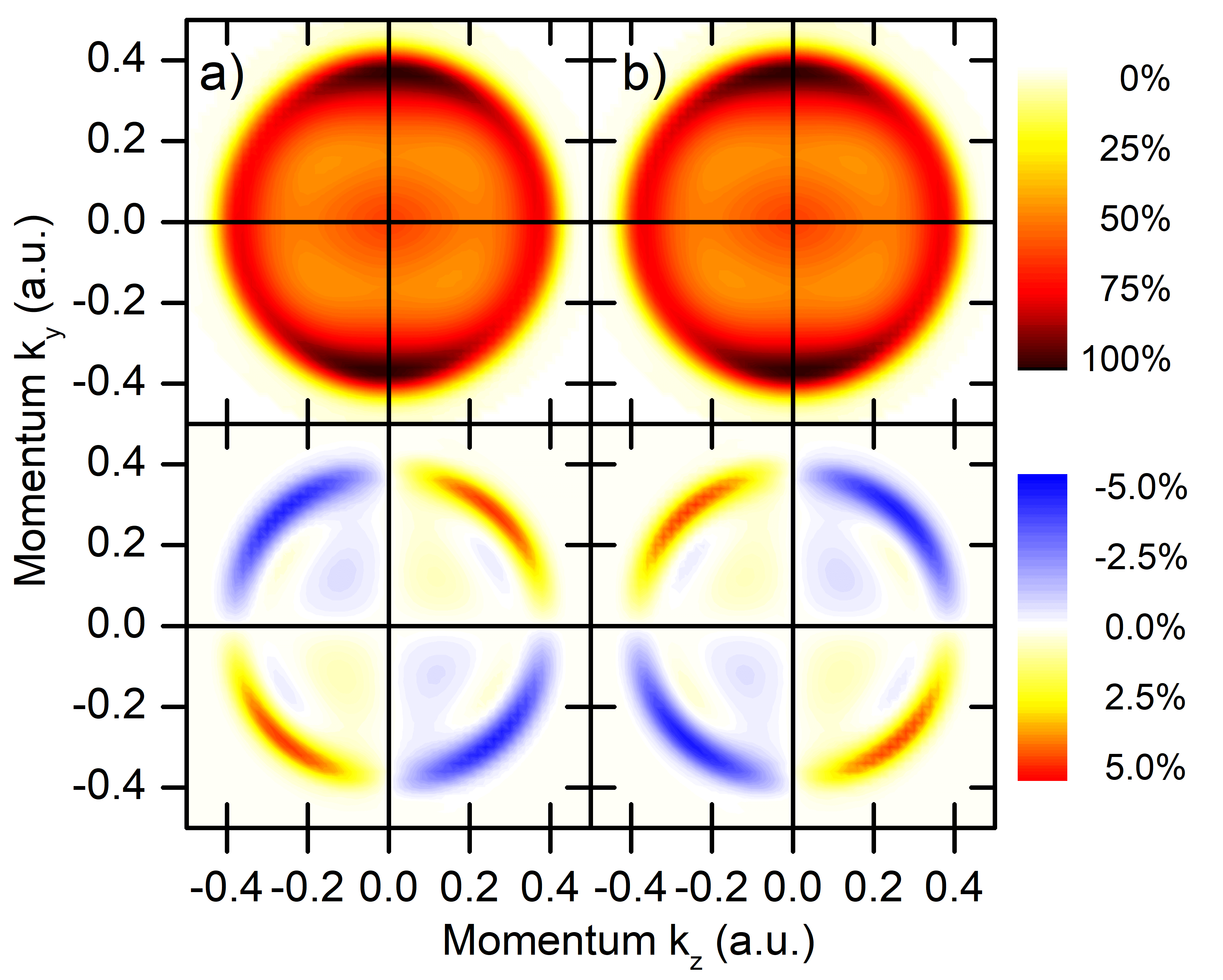}
\caption{Projections of the photoelectron angular distributions on the detector yz-plane (upper panels) and the respective PECDs (lower panels) computed for the case of six-red-photon vs. three-blue-photon ionization of the model chiral system by bichromatic pulses with $\omega=3.4$~eV and $2\omega=6.8$~eV  of equal peak intensities of $I_x=I_y=10^{13}$~W/cm$^2$ and pulse durations of $\tau=3$~fs. The observed chiral asymmetry is phase-sensitive: (a) $\phi=+\frac{\pi}{4}$ and (b)  $\phi=-\frac{\pi}{4}$. It is however somewhat smaller than that shown in Figs.~\ref{fig2}--\ref{fig4} for the two-red-photon vs. one-blue-photon ionization scheme.} \label{fig5}
\end{figure}

In conclusion, we demonstrate a possibility to study PECD phenomenon by a single experiment with two orthogonally-polarized laser pulses of frequencies $\omega$ and $2\omega$. Our simulations show a sizable forward/backward asymmetry, which depends on enantiomer and the rotational direction of the field. Nowadays, bichromatic pulses can be generated in many experimental laboratories in the optical \cite{Muller92,Yin92,Schumacher94,Yin95,Wang01,Yamazaki07,Vortex1,Vortex2,Vortex3,Gong14,Wu13,Dupont95,Hache97,Gudde07,XUV} and even XUV regimes \cite{FERMI}. The present results open a new avenue for future PECD experiments with bichromatic laser pulses. Particularly enticing are approaches in the optical regime making use of higher order multiphoton ionization schemes. Additional calculations performed in this work on the same model chiral system confirm the possibility to generate a notable effect by such schemes. As an example, results for the six-red-photon vs. three-blue-photon ionization scheme are displayed in Fig.~\ref{fig5}.

\begin{acknowledgements}
The authors thank T. Ring, C. Sarpe, H. Braun, A. Senftleben  and D. Reich for fruitful discussions. This work was supported by the Deutsche Forschungsgemeinschaft (DFG) within the Sonderforschungsbereich SFB--1319 `Extreme Light for Sensing and Driving Molecular Chirality -- ELCH' and the Schwerpunktprogramme SPP--1840 `Quantum Dynamics in Tailored Intense Fields -- QUTIF' (project 281051436).
\end{acknowledgements}

\end{document}